\documentclass[manuscript]{acmart}
\AtBeginDocument{%
  \providecommand\BibTeX{{%
    \normalfont B\kern-0.5em{\scshape i\kern-0.25em b}\kern-0.8em\TeX}}}

\acmConference[FAccTRec Workshop at RecSys '22]{Sixteenth ACM Conference on Recommender Systems}{September 18--23, 2022}{Seattle, WA, USA}
\setcopyright{none}
\usepackage{todonotes}
\usepackage{color}

\begin{document}

\title{RecSys Fairness Metrics: Many to Use But Which One To Choose?}

\author{Jessie J. Smith}
\email{jessie.smith-1@colorado.edu}
\affiliation{%
  \institution{University of Colorado Boulder}
  \country{USA}
}

\author{Lex Beattie}
\email{lex@spotify.com}
\affiliation{%
  \institution{Spotify}
  \country{USA}
}

\renewcommand{\shortauthors}{Smith et al.}

% \begin{abstract}
% No abstract required for this submission.
% \end{abstract}





\maketitle

\section{Introduction \& Background}

In recent years, recommendation and ranking systems have become increasingly popular on digital platforms. These often personalized systems leverage algorithms to recommend content, items, or information that matches users' perceived preferences. However, previous work has highlighted how personalized systems might lead to unintentional harms for users, such as degenerate feedback loops \cite{jiang2019degenerate, ribeiro2020auditing}, sexist stereotyping \cite{human2018only}, and racial bias \cite{julia-angwin}. Practitioners require metrics to measure and mitigate these types of harms in production systems. To meet this need, many \textbf{fairness definitions} have been introduced and explored by the RecSys community \cite{ekstrand2022fairness, deldjoo2022survey, kuhlman2021measuring, patro2022fair}. Unfortunately, this has led to a proliferation of possible \textbf{fairness metrics} from which practitioners can choose. The increase in volume and complexity of metrics creates a need for practitioners to deeply understand the nuances of fairness definitions and implementations. Additionally, practitioners need to understand the ethical guidelines that accompany these metrics for responsible implementation. \citet{jobin2019global} described the proliferation of ethics guidelines and found more than 80 documents containing ethical principles or guidelines for AI, pointing to the need for more implementation guidance rather than principles alone. The wide variety of available metrics, coupled with the lack of accepted standards or shared knowledge in practice \cite{richardson2021towards, deng2022exploring}, leads to a challenging environment for practitioners to navigate. In this position paper, we focus on this widening gap between the research community and practitioners concerning the availability of metrics versus the ability to put them into practice. We address this gap with our current work, which focuses on developing methods to help ML practitioners in their decision-making processes when picking fairness metrics for recommendation and ranking systems. In our iterative design interviews, we have already found that practitioners need both practical and reflective guidance when refining fairness constraints. This is especially salient given the growing challenge for practitioners to leverage the correct metrics while balancing complex fairness contexts.

\section{The Complexity of Fairness Metrics}
% Fortunately, many fairness definitions have been introduced and explored by the RecSys community \jessie{cite lit review papers}, and even more have been introduced broadly in the machine learning discipline \cite{narayanan2018translation, barocas-hardt-narayanan}. 

In machine learning, fairness definitions may have multiple associated fairness metrics. Additionally, each of these metrics may have unique associated parameters and thresholds that must be determined before a fairness measurement can occur. Measuring fairness in recommendation systems adds even more complexity to this space. For example, recommendation systems are often \textbf{multistakeholder} systems -- meaning they must cater to the needs of multiple groups of stakeholders \cite{burke2017multisided}. The two most common stakeholder groups are \textbf{providers} (those who provide or create content to be recommended) and \textbf{consumers} (those who interact with or consume the recommendations) \cite{burke2017multisided}. Fairness metrics can be used between or within each stakeholder group, and sometimes conflict with one another. Moreover, recommendation systems may consist of multiple components, meaning fairness needs to be measured within each component, from content generation and retrieval to pool re-ranking \cite{liu2019personalized, sonboli2020opportunistic, zehlike2017fa, sonboli2020fairness}. The combination of these variables for measuring fairness in recommendation systems compounds the complexity of choosing fairness metrics for practitioners. 

% For example, on a music recommendation platform like Spotify, we could measure if artists from certain regions of the world are consistently under-represented in recommendation lists, which might align with \textbf{provider fairness} goals. Or, we could measure if users from certain regions of the world consistently receive less accurate recommendations, which might align with \textbf{consumer fairness} goals. These two goals and their associated metrics might come into conflict with one another, and the decision for which goals and metrics to pursue are highly dependent on the organization's values and the context of the system.

Other decisions involved in choosing a metric include prioritizing between measuring group and individual fairness, determining quantifiable proxy variables for fairness, and defining qualitative fairness constraint(s). In machine learning fairness literature, researchers have broadly categorized fairness into two categories: group fairness versus individual fairness \cite{dwork2012fairness, binns2020apparent}. Group fairness measures if sensitive and/or non-sensitive groups acquire similar recommendation outcomes, while individual fairness requires that similar individuals are treated similarly. In recommendation and ranking, different metrics can measure group versus individual fairness within each stakeholder category \cite{ekstrand2022fairness}. Understanding how to differentiate between these fairness constraints and leveraging the correct metric for their context is one of the many complexities practitioners face.

% Even further, for group fairness, some metrics seek to measure fairness for one group at a time (e.g., is our protected provider group's distribution of recommendations equal to their target distribution?), while others seek to measure fairness in a binary setting (e.g., comparing ranking distribution of protected versus unprotected groups), or in a multi-group setting (e.g., what is the difference in distribution between providers from different countries in the world?) \cite{ekstrand2022fairness}.

In one paper, \citet{verma2020facets} classified RecSys fairness metrics as accuracy based, error based, and causal based. More recently, \citet{ekstrand2022fairness} published an in-depth review of fairness in recommendation systems. Unlike \citet{verma2020facets}, their review categorized pairwise fairness metrics with accuracy metrics, alleviating the potential confusion between distinguishing when a metric measures error versus accuracy. This difference in categorization reflects how fairness literature may change over time, making it difficult for practitioners to stay up to date and navigate this complex research space. In addition to understanding stakeholder and metric categories, practitioners must also understand how to implement their chosen metric correctly. Within each metric, various parameters and fairness thresholds can determine which fairness constraint the metric is attempting to measure. Leveraging different types of comparison distributions can cause practitioners to evaluate different fairness constraints \cite{geyik2019fairness}. Though it requires time and expertise to analyze all possible metric options, a team may feel rushed to `just choose one' for initial analysis to move towards an impactful audit. However, it can be difficult for practitioners to know if they are headed in the right direction.

\section{Choosing an Appropriate Fairness Metric}
In 2021, \citet{moss2021assembling} discussed \emph{Algorithmic Impact Assessments} to help engineers more easily report potential impacts of an algorithmic system. In practice, these assessments aim to help engineers describe potential impacts that their system might have on users in a worst-case scenario. These impact statements are perfect candidates for teams to map from a system's potential impact to a possible metric to quantify said impact. However, mapping from qualitative statements of values to quantitative proxies for measurement is no easy task. \citet{stray2022building} describe some of the difficulty in choosing an appropriate quantitative proxy for a qualitative construct in recommendation, a task previously defined as \textbf{construct validity} \cite{strauss2009construct}. In machine learning, construct validity can be challenging in all stages of deciding on a metric. These challenges include (1) determining if a plausible metric exists; (2) checking if the metric appropriately captures the qualitative constraints; (3) if the metric is comparable to other existing metrics; and (4) if the metric captures something different than previously used metrics \cite{jacobs2021measurement}.

Even without the challenge of achieving construct validity, practitioners encounter other obstacles when appropriately scoping fairness concerns for measurement. In one study, researchers discovered that ML practitioners have anxieties about their \emph{``blind spots''} when addressing fairness issues -- which could lead them to choose a metric that does not take certain vulnerable sub-populations into consideration \cite{holstein2019improving}. In another study, researchers discovered that some ML practitioners had similar anxiety around failing to identify the \emph{``correct''} fairness criteria for their users. For these participants, they mentioned that having additional resources about best practices for aligning fairness criteria with users' lived experiences would be beneficial for them to incorporate fairness into their existing ML workflows \cite{madaio2020co}. An alternate study reported that \emph{``participants told us that their organizations’ business imperatives dictated the resources available for their fairness work and that resources were made available only when business imperatives aligned with the need for disaggregated evaluations''} \cite{madaio2022assessing}. This difficulty arose partly because of the need for context-specific fairness metrics, which must account for the potentially competing interests of different user groups and the broader values of the organization hosting the AI system. However, the existence of broader organization values concerning fairness is not a given for practitioners \cite{cramer_practice}. Without organization-defined values and support, implementing contextual fairness in a large corporation could result in different teams leveraging competing fairness constraints or metrics, which could have potentially harmful downstream effects \cite{cramer_practice}. 

With so many metrics to choose from and metric implementation decisions to make, it can be daunting for a practitioner to attempt to measure and mitigate bias in their system. Additionally, most ML practitioners are not trained in disciplines like ethics or philosophy \cite{saltz2019integrating}, which creates another barrier to entry for deciding an appropriate fairness metric. Without institutional knowledge and academic knowledge of fairness metrics, choosing an appropriate metric might seem impossible. To combat some of these challenges, \citet{saleiro2018aequitas} created a decision tree for selecting an ML fairness metric. However, the decision tree assumes that the practitioner has prior knowledge of policy and ethics jargon, with some branches in the tree asking questions like, \emph{``are your interventions punitive or assistive.''} Additionally, this decision tree was designed for the context of binary classification, not ranking or recommendations. In the context of recommendation systems, fairness metrics and considerations are vastly different from a binary classification setting, especially since outcomes are not binary nor measurably favorable -- given that there is rarely a ``ground-truth'' to compare the final recommendation lists against beyond assuming user engagement as a positive prediction \cite{beutel_fairness_2019}. Thus, we see a gap in the RecSys discipline that needs to be addressed by helping ML practitioners decide on appropriate fairness metrics that complement their complex, contextual fairness considerations.

\section{Conclusion \& Future Work}
% ML practitioners have increased interest in measuring the impact that their systems have on their users, particularly in regards to fair and unfair treatment of certain sub-populations. 

Though academic literature has recently introduced dozens of fairness metrics, there are not enough resources that guide practitioners in choosing a metric that complements their specific contexts, organizational values, or prior knowledge -- especially in the discipline of recommendation and ranking. If fairness metrics only exist in academic papers, they will not be able to serve the purpose they were created for -- to help identify and measure unfair treatment or impact in machine learning systems. To promote and cultivate the use of fairness metrics in industry, we must start making tools and providing guidance to lower the barrier to entry for utilizing this knowledge. Our current research addresses this need. Through semi-structured interviews with real practitioners and an iterative design study, we have begun creating a decision-making framework to help practitioners choose a fairness metric that aligns with their specific fairness context. We have already begun uncovering specific challenges that practitioners face when refining fairness constraints or selecting a metric that matches their needs, and we are using this feedback to iterate on the design of a tool for alleviating these challenges. We recommend that future research should similarly include a strong focus on working with real practitioners and live recommendation systems in order to understand the real-world needs and obstacles that practitioners face when incorporating fairness into their pre-existing workflows and systems. Ideally, by collaborating with industry practitioners, we can enable fairness metrics to have real-world impact beyond the theoretical impact demonstrated via toy examples in academic papers. We hope this work can inspire the creation of tools, libraries, and guidelines to help practitioners evaluate fairness in a way that accurately captures the real experiences of their users and the practical constraints of online ranking and recommendation systems.

% We need interfaces for academic work to be applied in practice so that the field can learn from this, which includes increased collaboration between the two sectors.

% Thus, we propose a call to action for the RecSys research discipline to begin exploring what tools and guidance are most helpful for practitioners when deciding on a fairness metric for their given fairness context. 

% \begin{acks}

% \end{acks}

\bibliographystyle{ACM-Reference-Format}
\bibliography{bibliography}

% \appendix









\end{document}